\documentclass{article}

     \usepackage[final]{neurips_2019}
     \usepackage{ulem} 

\usepackage[utf8]{inputenc}
\usepackage{hyperref}
\usepackage{graphicx}
\usepackage{etoolbox}
\usepackage{bm}
\usepackage{bbm}
\usepackage{array}
\usepackage{amssymb} 
\usepackage{amsmath} 
\usepackage{fancyhdr} 
\usepackage{amsthm} 
\usepackage{mathtools} 

\usepackage{qcircuit} 
\usepackage{algorithm}
\usepackage{algpseudocode}

\usepackage{tikz}

\graphicspath{{./}}

\newcommand\mapsfrom{\mathrel{\reflectbox{\ensuremath{\mapsto}}}}

\DeclareMathOperator*{\argmax}{argmax}
\DeclareMathOperator*{\argmin}{argmin}

\def\E{\mathbb E}
\def\P{\mathbb P}

\def\R{\mathbb R}
\def\Z{\mathbb Z}

\def\A{\mathcal{A}} 

\def\One{\mathbbm 1} 

\def\RR{\mathcal R}
\def\LL{\mathcal{L}}
\def\C{\mathbb C}

\usepackage{hyperref}       

\hypersetup{ 
    colorlinks=true,
    citecolor=blue,
    linkcolor=blue,
    filecolor=magenta,      
    urlcolor=blue,
}

\title{Global Optimum Search in Quantum Deep Learning}

\author{%
  Lanston Hau Man Chu \\
  University of Wisconsin-Madison\\
  Madison, WI 53706 \\
  \texttt{hchu34@wisc.edu} \\
  \And
  Tejas Bhojraj \\
  University of Wisconsin-Madison\\
  Madison, WI 53706 \\
  \texttt{bhojraj@wisc.edu} \\
  \AND
  Rui Huang\\
  University of Wisconsin-Madison\\
  Madison, WI 53706 \\
  \texttt{huangrui@cs.wisc.edu} \\
}

\begin{document}

\maketitle

\begin{abstract}
This paper aims to solve machine learning optimization problem by using quantum circuit. Two approaches, namely the average approach and the Partial Swap Test Cut-off method (PSTC) was proposed to search for the global minimum/maximum of two different objective functions. The current cost is $O(\sqrt{|\Theta|} N)$, but there is potential to improve PSTC further to $O(\sqrt{|\Theta|} \cdot sublinear \ N)$ by enhancing the checking process.

\end{abstract}

\section{Introduction}

Recent years have witnessed great success of machine learning in a wide range of applications, such as computer vision, natural language processing, speech recognition and so on. AlphaGo, a product of reinforcement learning, has beaten the human world champion in Go. A daily life example of machine learning algorithm is Amazon's recommendations system. Moreover, machine learning has also found its applications in many other science fields, such as physics, communication engineering, material science, economics and so on.

Supervised learning is one of the most popular areas in machine learning. A large number of this type of machine learning tasks, especially those involving deep neural networks, rely on an optimization problem formulated as below:

\begin{equation}
\label{eq:optimization_problem_without_mentioning_average_or_cutoff}
\theta^{*} = \argmin_{\theta \in \Theta} \LL(\theta)
\end{equation}

where $\theta$ is the model parameter and $\LL(\cdot)$ is the corresponding objective function (often to be the empirical expected loss $\frac{1}{N} \sum\limits_{j=1}^N loss_{\theta}(x_j) = \frac{1}{N} \sum\limits_{j=1}^N \ell(\theta, x_j)$. The most prominent classic algorithm to solve (\ref{eq:optimization_problem_without_mentioning_average_or_cutoff}) is the gradient descent (GD) algorithm. The key idea behind gradient descent is to iteratively move $\theta$ towards the direction of steepest descent until we reach a convergence (as shown in Equation~\ref{eq:gd}).


\begin{equation}
\label{eq:gd}
\theta_{k+1} = \theta_k - \eta \frac{\partial \LL_\theta}{\partial \theta} \Big|_{\theta = \theta_k}
\end{equation}

where $\LL_\theta$ is the empirical expected loss to be maximized and $\theta_k$ is the value of $\theta$ estimated at iteration $k$. Even though GD has achieved empirical successes in numerous machine learning models and applications, there are some drawbacks in the following two aspects. First, depending on the initial point, GD may find a local optimum instead of the global optimum. In that case, the algorithm will stuck at a local optimum and there is no way to tell whether we are converging to a local or global optimum. Second, the upper bound or expected number of iterations of (\ref{eq:gd}) to reach convergence is difficult to determine especially for complicated deep learning model architecture.

In contrast, in this paper we propose two quantum optimization algorithms (namely, the average approach and the cut-off approach PSTC) to search for the global optimum for (\ref{eq:optimization_problem_without_mentioning_average_or_cutoff}) without using gradient descent. Our algorithms are inspired by the Grover's search \cite{grover1996fast} (for specific value searching), the Durr \& Hoyer algorithm \cite{Durr1999} (DH algorithm; for optimum value searching), as well as the swap test \cite{Swap_Test}.

Our first approach in section \ref{section:approach_1} (i.e. the ``average approach'') uses DH algorithm to minimize the traditional objective function. Our second approach in section \ref{section:approach_2} is named as the \textbf{Partial Swap Test Cut-off method} (i.e. PSTC; simply refer as the ``cut-off'' approach in our paper), in which we propose a modified objective function to maximize the number of ``cut-off indicators $\E_{\theta j}$'' (indicators on whether loss is below a certain threshold) instead of minimizing the loss average over all training items. Then we utilize a ``partial'' swap test and amplitude amplification $\A_{\text{Boost}}$ to keep updating the model parameter, until we find the optimal parameter $\theta$. Both two approaches are aimed to search for the global optimum with the expected cost of $O(\sqrt{|\Theta|} N)$, where $N$ is the number of total training samples and $|\Theta|$ is the size of parameter space. Our contributions can be summarized as follows:

\begin{itemize}
    \item We propose two quantum approaches (i.e. the average approach and the cut-off approach PSTC) to find the global optimum (instead of a local optimum) for optimizing machine learning models.
    \item We provide theoretical analyses for both approaches to show that the expected cost are both $O(\sqrt{|\Theta|} N)$.
    \item We design a novel objective function (\ref{eq:quantum_parallelism_PSTC}) maximizing the number of ``cut-off indicators $\E_{\theta j}$'' to fit the property of quantum computing in optimizing machine learning models
    \item While approach 1 and 2 have the same cost $O(\sqrt{|\Theta|} N)$, we will show that the cost $O(N)$ of average approach (approach 1) is mainly at the quantum parallelism part, and the $O(N)$ of PSTC (approach 2) is mainly at the checking process. Therefore there is potential to improve PSTC to reduce the cost from $O(\sqrt{|\Theta|} N)$ to $O(\sqrt{|\Theta|} \cdot sublinear \ N)$ in future work
\end{itemize}

This paper will be organized as below: We will compare the theoretical framework of the two approaches (section \ref{section:overview}), and then introduce our average approach (section \ref{section:approach_1}) and cut-off approach (PSTC; section \ref{section:approach_2}) respectively. We will discuss the relationship between the objective functions of the two approaches (section \ref{section:equivalency_discussion}), and make conclusion (section \ref{section:conclusion}). We will talk about future work and extension (section \ref{section:future_work_extension}), and there is a notations table (section \ref{section:notations_table}) and appendix with proofs (section \ref{section:appendix}) at the end of the paper.

\section{Overview and Theory}
\label{section:overview}

In this paper, we propose two quantum optimization algorithms (i.e. the average approach, and the cut-off approach PSTC) without using gradient descent to solve optimization problem (\ref{eq:optimization_problem_without_mentioning_average_or_cutoff}).

The average approach (approach 1) utilizes DH algorithm to find the global minimum with the expected cost being $O(N\sqrt{\Theta})$. We treat the machine learning model that calculates loss for each sample as our quantum black box:

\begin{equation}
\label{eq:quantum_parallelism_avg}
\text{Average approach: } \begin{cases} | \theta \rangle | j \rangle | \textbf{0} \rangle \xmapsto[]{\text{Quantum parallelism with cost } O(N)} | \theta \rangle | j \rangle | \sum\limits_{j = 1}^N \ell(\theta, x_j) \rangle \\ \theta^{*}_{avg} = \argmin\limits_{\theta \in \Theta} \LL^{avg}_\theta = \argmin\limits_{\theta \in \Theta} \frac{1}{N} \sum\limits_{j=1}^N \ell(\theta, x_j) \end{cases}
\end{equation}

where $\theta \in \Theta$ is a model parameter and $x_j (j \in \{1,2,\dots, N\})$ is a training sample, and we want to minimize the function $\sum_j \ell(\theta, x_j)$ by searching over $\theta$. In each step, we use classical approach to sum over the total loss over all training items. Therefore, the cost of each step is $O(N)$, where $N$ is the number of training items. Note that, in the first approach, $O(N)$ queries are fundamental at each step and the potential of improving this cost is limited.

To mitigate this cost, we propose the cut-off approach (PSTC) as our second approach, where we focus on a cut-off indicator $E_{\theta j} \triangleq \One[\, \ell(\theta, x_j) \leq \ell_{threshold}]\,$ with a revised objective function $\LL^{\text{PSTC}}$:

\begin{small}
\begin{equation}
\label{eq:quantum_parallelism_PSTC}
\text{Cut-off approach (PSTC): } \begin{cases} | \theta \rangle | j \rangle | \textbf{0} \rangle \xmapsto[]{\text{Quantum parallelism with cost } O(1)} | \theta \rangle | j \rangle | E_{\theta j} \rangle \\ \theta^{*}_{\text{PSTC}} = \argmax\limits_{\theta \in \Theta} \LL^{\text{PSTC}}_\theta = \argmax\limits_{\theta \in \Theta} \frac{1}{N} \sum\limits_{j=1}^N \One[\,\ell(\theta, x_j) \leq \ell_{threshold}]\, \end{cases}
\end{equation}
\end{small}

As we can see, we had proposed a novel objective function for PSTC which is different from the traditional one. Instead of minimizing the sum of loss over all training items, we maximize the number of training items with loss below a certain threshold.

In PSTC, we still need the cost $O(N)$ queries (to be shown in section \ref{section:potential_cost_improvement}) to check whether the observed $\theta$ is ``good'' (i.e. with high $\LL^{\text{PSTC}}_\theta$). We have therefore shifted the burden of cost $O(N)$ from quantum parallelism to the checking process. It is worth to point out that overall cost of the average approach and PSTC is still same as $O(\sqrt{|\Theta|} N)$ as of writing of this paper. But we believe that there are room for PSTC to improve further by using some dynamic trick on the checking process, since the cost at quantum parallelism is $O(1)$ only.

In order to boost the probability of finding a ``good'' $\theta_{\text{new}}$, we employ a ``partial" swap test and amplitude amplification in each iteration.

Since in approach 1 and approach 2 we are using different objective function $\LL^{avg}$ and $\LL^{\text{PSTC}}$, it is also worth to discuss whether this two different objective function are `` equivalent'' in some sense, which please refer to section \ref{section:equivalency_discussion}.

\section{Approach 1: Average Approach}
\label{section:approach_1} 

Let $N=2^n$ be the sample size and $x_j$ denote the samples of the sample set $S = \{ x_1, ..., x_N\}$ and $i$ is the index of the parameters $\{ \theta_i \}_{i=1}^M$. Let $|\Theta|=M=2^m$ be the number of parameters. Let $t$ be the value such that $f(i)\leq 2^t$, where $f$ is to be defined below.

Our algorithm is as follows:
\begin{enumerate}
    \item Consider the function $f(i)=f(\theta_i):= \sum_j \ell (\theta_i, x_j)$
    \item Use the Durr-Hoyer algorithm \cite{Durr1999} to find the minimum of $f$ with a slight modification: When the DH algorithm queries $f(i)$ for some $i$ (i.e. when the algorithm makes a query on $|i\rangle |0^t\rangle$ and needs $|i\rangle |f(i)\rangle$ as output), do the following subroutine: 
    \begin{itemize}
        \item Classically compute $f(i)$
        \item Build a unitary operator, $U$ mapping $|i\rangle |0^t\rangle \mapsto |i\rangle |f(i)\rangle$.
        \item Return $U(|i\rangle |0^t\rangle)$ to the DH algorithm.
    \end{itemize}

\end{enumerate} 

\emph{Analysis:}
Each time we invoke the subroutine, we need $N$ classical queries to $\ell$. Also, the DH algorithm makes an expected number of $O(\sqrt{|\Theta|})$ queries. So, we run the subroutine $O(\sqrt{|\Theta|})$ times on average. So, the cost is $O(\sqrt{|\Theta|} N)$. Also, we know that the DH algorithm succeeds with probability $0.5$ and so by running it many times, we get the parameter minimizing the loss with arbitrarily good accuracy at the cost $O(\sqrt{|\Theta|} N)$.

\section{Approach 2: ``Cut-off'' Approach (PSTC)}
\label{section:approach_2}

There are four algorithms being used in the \textbf{Partial Swap Test Cut-off method} (PSTC), namely $\A_{\text{1-query}}$, $\A_{\xi}$, $\A_{\text{Boost}}$ and $\A_{\text{PSTC}}$. The key routine is $\A_{\text{1-query}}$ while the main algorithm to be run is $\A_{\text{PSTC}}$. This section still start from the introduction of the cut-off indicator $E_{\theta j}$ for the quantum circuit $Q_{\text{1-query}}$ of $\A_{\text{1-query}}$.

\subsection{The cut-off indicator $E_{\theta j}$ and loss threshold}

In the average approach (i.e. approach 1), we are doing the quantum parallelism of $| \theta _i \rangle | 0 \rangle \mapsto | \theta _i \rangle | \sum\limits_j \ell _{ij} \rangle$ in (\ref{eq:quantum_parallelism_avg}) which requires cost $O(N)$ for the summation. We want to improve the algorithm by moving the summation sign out of the ket-notation to avoid such a high cost. In other words, we are going to move the $O(N)$ from the quantum parallelism parts to other parts (i.e. to the checking process, which we will show later on). After doing this in PSTC, summation sign of the quantum parallelism will be outside the ket-notation, and the cost for quantum parallelism is changed from $O(N)$ to $O(1)$.

In PSTC (i.e. approach 2, the cut-off approach), we will pick a threshold value $\tilde{\ell}$ (or written as $\ell_{threshold}$) of loss. For example, $\tilde{\ell}$ can be chosen by random sampling of $x$ and take the average of the loss of the samples. We will then obtain a 1-qubit indicator $E_{\theta j}$. Instead of the original average loss objective function $\LL^{avg}_\theta$, we will use the alternative loss objective function $\LL^{\text{PSTC}}_\theta$:

\begin{equation}
\label{eq:L_cut_off}
\LL^{\text{PSTC}}_\theta \triangleq \frac{1}{N} \sum\limits_{j=1}^N E_{\theta j} \text{, where } E_{\theta j} = \One[\, \ell(\theta, x_j) \leq \tilde{\ell} ]\,
\end{equation}

Recall that $N$ is the size for our sample set $S = \{ x_1, ..., x_N\}$. Note that $E_{\theta j}$ and $E_{i j}$ can be written interchangeably especially when $\theta = \theta_i$. We are going to do the quantum parallelism as below:

\[ \sum\limits_{\theta \in \Theta} \sum\limits_{j=1}^N | \theta \rangle | j \rangle | \textbf{0} \rangle \xmapsto[]{\text{quantum parallelism with cost } O(1)} \sum\limits_{\theta \in \Theta} \sum\limits_{j=1}^N | \theta \rangle | j \rangle | E_{\theta j} \rangle\]

The summation sign $\sum\limits_{j=1}^N$ over different samples $x$ is now outside the ket-notation, which reduce the cost of quantum parallelism from $O(N)$ to $O(1)$. Since we are doing optimization on the cut-off indicator $E_{\theta j}$ instead of the loss $\ell_{\theta j} \triangleq \ell(\theta, x_j)$ per se, we call our new approach (i.e. approach 2) the \textbf{cut-off approach}. Once we talked about the details of the algorithm, we will show that the partial swap test will also be involved, and therefore we will formally name our methodology as the \textbf{Partial Swap Test Cut-off method} (PSTC), or simply the cut-off approach without causing any confusion in our paper. Please note that the sum of qubits $E_{\theta j}$ have the following meaning for the model parameter $\theta$:

\[ \sum\limits_{j = 1}^N E_{\theta j} = \#( \text{elements in } S \text{ that have } loss_{\theta} \leq \tilde{\ell}) \]

which is the number of elements in $S$ that have loss smaller than the chosen threshold.

In contrast to $\LL^{\text{avg}}_\theta$ which is the smaller the better, now \textbf{$\LL^{\text{PSTC}}_\theta$ is the larger the better}, i.e. large $\LL^{\text{PSTC}}_\theta$ gives us low $\ell$ in general.

In this paper we will also discuss whether $\LL^{\text{avg}}_\theta$ and $\LL^{\text{PSTC}}_\theta$ can be considered as ``equivalent'' in some sense, which please refer to section \ref{section:equivalency_discussion}.

\subsection{Partial swap test}

To compute $\LL^{\text{PSTC}}_\theta$ in (\ref{eq:L_cut_off}), we will use the idea of swap test \cite{Swap_Test} for the inner product. Instead of using the original swap test, the swap test is now only treated as a partial circuit of our sub-routine $Q_{\text{1-query}}$, and we will measure $| \theta \rangle$ to obtain some ``good'' $\theta$ with low loss $\ell$. We call this approach the \textbf{Partial Swap Test}.

\subsubsection{Relationship of $\LL^{\text{PSTC}}_\theta$ and the inner product of states}

First of all, we now define and construct two $(n+1)$-qubit state as below, where $N = 2^n$ is the number of samples $x$ in $S = \{ x_1, ..., x_N\}$:

\begin{equation}
\label{eq:phi_and_psi}
\begin{cases} | \phi_\theta \rangle \triangleq \frac{1}{\sqrt{N}} \sum\limits_{j = 1}^N | j \rangle | E_{\theta j} \rangle \\ | \psi \rangle \triangleq \frac{1}{\sqrt{N}} \sum\limits_{k = 1}^N | k \rangle | 1 \rangle \end{cases}
\end{equation}

Note that $| \phi_\theta \rangle$ is a function of $\theta$, while $| \psi \rangle$ is independent to $\theta$. We can see that the inner product of the two states is in fact $\LL^{\text{PSTC}}_\theta$:

\[ \langle \phi_\theta | \psi \rangle = \Big( \frac{1}{\sqrt{N}} \sum\limits_{j = 1}^N \langle j | \langle E_{\theta j} | \Big) \Big( \frac{1}{\sqrt{N}} \sum\limits_{k = 1}^N | k \rangle | 1 \rangle \Big) = \frac{1}{N} \sum\limits_{j = 1}^N \langle j | j \rangle E_{\theta j} = \LL^{\text{PSTC}}_\theta\]

The second equal sign is achieved by $\langle j | k \rangle = 0$ for $j \neq k$ and $E_{\theta j} \in \{ 0 ,1\} \implies \langle E_{\theta j} | 1 \rangle = E_{\theta j}$, which is mainly the reason why we want $| E_{\theta j} \rangle$ to be 1-qubit only.

As mentioned above, now $\LL^{\text{PSTC}}_\theta$ is the larger the better, i.e. large $\LL^{\text{PSTC}}_\theta$ gives us low $\ell$ in general. To get a large loss $\LL^{\text{PSTC}}_\theta$, we want $\langle \phi_\theta | \psi \rangle$ to be large for some $\theta$, i.e.

\begin{equation}
\label{eq:max_L_cut_off}
\boxed{
\theta^{*}_{\text{PSTC}} = \argmax\limits_{\theta \in \Theta} \LL^{\text{PSTC}}_\theta = \argmax\limits_{\theta \in \Theta} \langle \phi_\theta | \psi \rangle = \argmax\limits_{\theta \in \Theta} \frac{1}{N} \sum\limits_{j=1} E_{\theta j}
}
\end{equation}

Though practically that is difficult to get the best $\theta^{*}$, we want to obtain some ``good'' $\theta$ which gives us some large value of $\LL^{\text{PSTC}}_\theta$.

\subsubsection{The Q-circuit $Q_{\text{1-query}}$}

The inner product $\langle \phi_\theta | \psi \rangle = \frac{1}{N} \sum\limits_{j = 1}^N E_{\theta j} = \LL^{\text{PSTC}}_\theta$ is the key to find ``good'' $\theta$ with large $\LL^{\text{PSTC}}_\theta$.

To achieve this, we introduce the quantum circuit (i.e. Q-circuit) $Q_{\text{1-query}}$ for the computation:

\begin{figure}[ht]
\caption{$Q_{\text{1-query}}$: The Partial Swap Test circuit}
\label{fig:Q_1_query}
\begin{equation}
\Qcircuit @C=1em @R=2em {
\lstick{| 0 \rangle} & {/ ^1} \qw & \gate{H} & \ctrl{3} & \gate{H} \ar@{.}[]+<1.2em,1em>;[d]+<1.2em,-8.7em> & \meter & \ctrl{4} & \qw \\
\lstick{\frac{1}{\sqrt{|\Theta|}} \sum\limits_{\theta \in \Theta} | \phi_\theta \rangle = \frac{1}{\sqrt{|\Theta|}} \sum\limits_{\theta \in \Theta} \Big( \frac{1}{\sqrt{N}} \sum\limits_{j = 1}^N | j \rangle | E_{\theta j} \rangle \Big) } & {/ ^{n+1}} \qw & \qw & \qswap & \qw & \qw & \qw & \qw \\
\lstick{} &  &  &  &  &  & & \\
\lstick{| \psi \rangle = \frac{1}{\sqrt{N}} \sum\limits_{k = 1}^N | k \rangle | 1 \rangle} & {/ ^{n+1}} \qw & \qw & \qswap & \qw & \qw & \qw & \qw \\
\lstick{\frac{1}{\sqrt{|\Theta|}} \sum\limits_{\theta \in \Theta} | \theta \rangle} & \qw & \qw & \qw & \qw & \qw & \meter & \qw \\
& & & & \quad \quad \ \ | \varsigma_\theta \rangle & & & \\
}
\end{equation}
\end{figure}
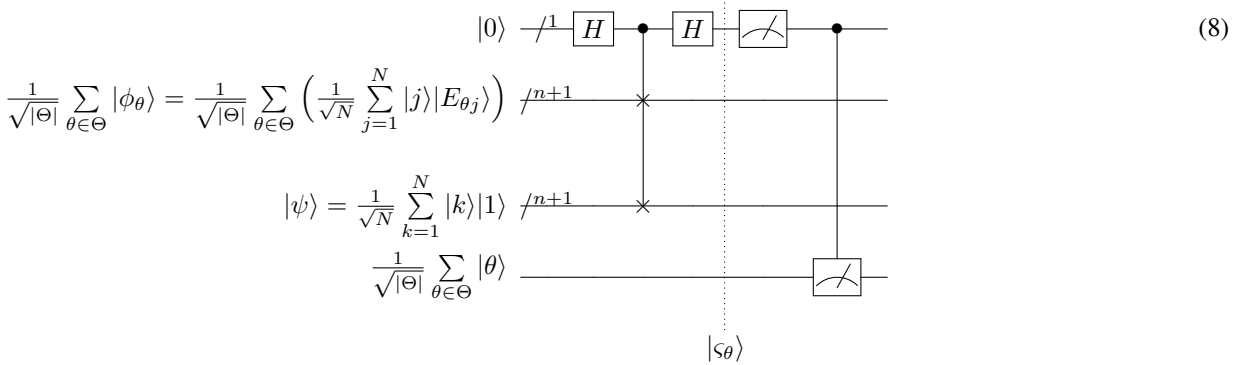

After the 2nd H-gate, if we treat $\theta$ as a fixed value, that is a standard swap test \cite{Swap_Test} and the state at the dashed line would be (ignoring the last wire of $| \theta \rangle$, as well as $\sum\limits_{\theta \in \Theta}$):

\[ | \varsigma_\theta \rangle = \frac{1}{2} | 0 \rangle \Big( | \phi_\theta \rangle | \psi \rangle + | \psi \rangle | \phi_\theta \rangle \Big) + \frac{1}{2} | 1 \rangle \Big( | \phi_\theta \rangle | \psi \rangle - | \psi \rangle | \phi_\theta \rangle \Big) \]

If we consider $\frac{1}{\sqrt{|\Theta|}} \sum\limits_{\theta \in \Theta} | \theta \rangle$ instead of one $| \theta \rangle$ (which should not be difficult to achieve by applying H-gate before $Q_{\text{1-query}}$), the state at the dashed line would be (i.e. Lemma 2.0 of section \ref{section:lemma_2_0}):

\[ \frac{1}{\sqrt{|\Theta|}} \sum\limits_{\theta \in \Theta} | \varsigma_\theta \rangle | \theta \rangle = \frac{1}{\sqrt{|\Theta|}} \sum\limits_{\theta \in \Theta} \frac{1}{2} | 0 \rangle \Big( | \phi_\theta \rangle | \psi \rangle + | \psi \rangle | \phi_\theta \rangle \Big) | \theta \rangle + \frac{1}{2} | 1 \rangle \Big( | \phi_\theta \rangle | \psi \rangle - | \psi \rangle | \phi_\theta \rangle \Big) | \theta \rangle \]

This gives us the following result (please refer to sections \ref{section:lemma_2_1}, \ref{section:lemma_2_2}, \ref{section:theorem_2_4} of \textbf{Appendix} for the proof:)

\[
\implies \begin{cases}
\text{Lemma 2.1: } \P \Big[ \text{1st qubit} = 0 | \theta \Big] = \frac{1}{2} + \frac{1}{2} |\langle \phi_\theta, \psi \rangle |^2 \\
\text{Lemma 2.2: } \P \Big[ \text{1st qubit} = 0 \Big] = \frac{1}{2} + \frac{1}{2 | \Theta |} \sum\limits_{\theta \in \Theta} |\langle \phi_\theta, \psi \rangle |^2
\end{cases}
\]

\[ \implies \boxed{\text{Theorem 2.4: } \P \Big[ \text{observe } \theta | \text{1st qubit = 0} \Big] \propto \frac{1}{2} + \frac{1}{2} \Big| \frac{1}{N} \sum\limits_{j = 1}^N E_{\theta j} \Big|^2 } \quad = \frac{1}{2} + \frac{1}{2} \Big| \LL^{\text{PSTC}}_\theta \Big|^2 \]

In (\ref{eq:max_L_cut_off}), we want $\theta$ with large $\LL^{\text{PSTC}}_\theta$, and \textbf{Theorem 2.4} tells us that such good $\theta$ can be observed with higher probability than other bad $\theta$. In other words, the \textbf{better $\theta$ we want, the higher chance we can observe}! Unlike the original swap test which do several measurement to estimate $\langle \phi_\theta | \psi \rangle$, we only do the observation once in each iteration of $\A_{\text{1-query}}$ to get a ``good'' $\theta$ according to the probability distribution based on the value of $\langle \phi_\theta | \psi \rangle$. Besides, the original swap test circuit is just a partial circuit of $Q_{\text{1-query}}$ while consider one possible $| \theta \rangle$ only. We call the entire $Q_{\text{1-query}}$ the \textbf{Partial Swap Test}, which will be used in our algorithm $\A_{\text{1-query}}$, and the algorithm is now listed as below:

\begin{algorithm}
  \caption{$\A_{\text{1-query}}:$ 1-query routine to search for a good $\theta$}
  \begin{algorithmic}[1]
      \State \% \textbf{Main idea}: Subroutine to find a good $\theta$ with small loss
      \State \% \textbf{Input}: $\begin{cases} \Theta, \quad \xi_{threshold}, \quad \ell_{threshold} \\ \{ \sum\limits_{\theta \in \Theta} | E_{\theta j} \rangle : x_j \in S\} \\ \sum\limits_{\theta \in \Theta} | \theta \rangle, \quad \sum\limits_{j = 1}^N | j \rangle, \quad \sum\limits_{k = 1}^N | k \rangle \end{cases}$
      \State \% \textbf{Sub-algorithm}: $Q_{\text{1-query}}$
      \State \% \textbf{Output}: $(\theta_{\text{1-query}}, flag)$ \Comment{$flag$ tells whether $\theta_{\text{1-query}}$ is good}
      \State \% \textbf{Analysis}: $\P[\, \text{observe } \theta | \text{1st qubit} = 0]\, \propto \frac{1}{2} + \frac{1}{2} \Big| \frac{1}{N} \sum\limits_{j = 1}^N E_{\theta j} \Big|^2$
      \State \% \textbf{Complexity}: $O(N)$
      \State Run $Q_{\text{1-query}}$ \Comment{Complexity of $Q_{\text{1-query}}: O(1)$}
      \State Do \Qcircuit @C=0em @R=2em {\lstick{} & \meter} $\;$ on the 1st qubit
		\If{1st qubit $= $ `` $1$''} \Comment{Doesn't meet requirement to proceed}
		\State Restart $\A_{\text{1-query}}$
		\ElsIf{1st qubit $= $ `` $0$''} \Comment{Meet requirement to proceed}
		\State Do \Qcircuit @C=0em @R=2em {\lstick{} & \meter} $\;$ on $| \theta \rangle$ to obtain $\theta_{\text{1-query}}$
        \EndIf
        \State \% The \textbf{Checking Process} \Comment{Complexity of $A_\xi: O(N)$}
        \State $\xi(\theta_{\text{1-query}}) \mapsfrom$ Run $\A_{\xi}(\theta_{\text{1-query}})$ \Comment{Check whether $\theta_{\text{1-query}}$ is within region $R$;}
		\If{$\xi(\theta_{\text{1-query}}) < \xi_{threshold} $}
		\State $flag \mapsfrom 0$ \Comment{Bad $\theta \notin R$}
		\ElsIf{$\xi(\theta_{\text{1-query}}) \geq \xi_{threshold} $}
		\State $flag \mapsfrom 1$ \Comment{Good $\theta \in R$}
		\EndIf
        \State \textbf{Return} $(\theta_{\text{1-query}}, flag)$
  \end{algorithmic}

\end{algorithm}

For sub-algorithm $\A_\xi$, region $R$ and cost $O(N)$ mentioned in the checking process of $\A_{\text{1-query}}$, they will be elaborated further when we talked about $\A_{\text{Boost}}$.

It is also worth to point out that Lemma 2.2 (i.e. section \ref{section:lemma_2_2}) tells us the cost of getting a zero for the first qubit is $1/\P \Big[ \text{1st qubit} = 0 \Big] = O\Big( \Big( \frac{1}{2} + \frac{1}{2} \Big| \frac{1}{N} \sum\limits_{j = 1}^N E_{\theta j} \Big|^2 \Big)^{-1} \Big) = O(1)$ by intuition given that we have some proper choice of $\ell_{threshold}$, and therefore such cost won't be too large.

\subsection{Boosting: The amplitude amplification}

Now we define $\xi(\theta)$ as below and will compute it in the algorithm $\A_\xi$ (refer to Appendix) with cost $O(N)$ below:

\[ \xi(\theta) \triangleq \frac{1}{2} + \frac{1}{2} \Big| \frac{1}{N} \sum\limits_{j = 1}^N E_{\theta j} \Big|^2 \]

\begin{algorithm}
  \caption{$\A_{\xi}:$ Function to estimate $\P[\, \text{observe } \theta | \text{1st qubit} = 0]\,$}
  \begin{algorithmic}[1]
      \State \% \textbf{Main idea}: Classic summation to compute $\xi(\theta) \triangleq \frac{1}{2} + \frac{1}{2} \Big| \frac{1}{N} \sum\limits_{j = 1}^N E_{\theta j} \Big|^2$
      \State \% \textbf{Input}: $\begin{cases} \theta \\ S = \{ x_1, x_2, ..., x_N\} \\ \ell_{threshold} \end{cases}$
      \State \% \textbf{Sub-algorithm}: Loss function $\ell(\cdot, \cdot)$
      \State \% \textbf{Output}: $\xi(\theta) \in \R$
      \State \% \textbf{Analysis}: Nil
      \State \% \textbf{Complexity}: $O(N)$
      \State $\ell_{\theta j} \mapsfrom \ell(\theta, x_j)$
      \State $ E_{\theta j} \mapsfrom \One[\, \ell_{\theta j} \leq \ell_{threshold}]\,$
      \State $ \xi(\theta) \mapsfrom \frac{1}{2} + \frac{1}{2} \Big| \frac{1}{N} \sum\limits_{j = 1}^N E_{\theta j} \Big|^2$
      \State \textbf{Return} $\xi(\theta)$
  \end{algorithmic}
\end{algorithm}

Luckily, while we want to maximize $\xi(\theta)$ as a maximization of (\ref{eq:max_L_cut_off}), by \textbf{Theorem 2.4} we can also see that $\xi(\theta) \propto \P \Big[ \text{observe } \theta | \text{1st qubit = 0} \Big]$, which is the probability mass function (PMF) of $\theta$. Therefore, we can choose some threshold value $\tilde{\xi}$ (or written as $\xi_{threshold}$), such that for any $\theta$ with $\xi(\theta) \geq \tilde{\xi}$, we will say $\theta$ is in the region $R$ (i.e. $\theta \in R \subseteq \Theta$). Now the goodness of $\theta$ is well defined: We will say $\theta$ is a ``good'' $\theta \iff \theta \in R$. The picture of $\text{PMF}(\theta)$ is referring to table \ref{table:PMF_xi_theta} as below:

\begin{table}[ht]
\caption{Amplitude amplification for the $\text{PMF}(\theta) \propto \xi(\theta)$}
\label{table:PMF_xi_theta}
\begin{center}
\begin{tabular}{c c}

\resizebox{5cm}{!}{
\begin{tikzpicture}
\def\normaltwo{\x,{-1 + 4*1/exp(((\x-3)^2)/(2*6))}}

\newcommand{\xiValue}[1]{%
-1 + 4*1/exp(((#1-3)^2)/(2*6))
}

\fill [fill=orange!60] (1.6,0) -- plot[domain=1.6:4.4] (\normaltwo) -- (4.4,0) -- cycle;

\draw[color=blue,domain=0:6] plot (\normaltwo) node[right] {};

\draw[dashed] (4.4,{\xiValue{4.4}}) -- (4.4,0);
\draw[dashed] (1.6,{\xiValue{1.6}}) -- (1.6,0);
\draw[dashed] (1.6,{\xiValue{1.6}}) -- (0,{\xiValue{1.6}}) node[left] {$\xi_{threshold}$};

\draw (1.5,5) node[above] {$\text{PMF}(\theta) \propto \xi(\theta) = \frac{1}{2} + \frac{1}{2} \Big| \frac{1}{N} \sum\limits_{j = 1}^N E_{\theta j} \Big|^2 $};
\draw (6.2,0) node[right] {$\theta \in \Theta$};

\draw[->] (0,0) -- (6.2,0) node[right] {};
\draw[->] (0,0) -- (0,5) node[above] {};
\filldraw[fill=green, draw=green!40!black] (1.6,0.3) rectangle (4.4,-0.2);
\draw (3,0) node {Region $R$};

\draw[color=purple] (1.6,-0.5) -- (4.4,-0.5);
\draw[color=purple] (1.6,-0.7) -- (1.6,-0.3);
\draw[color=purple] (4.4,-0.7) -- (4.4,-0.3);
\draw[color=purple] (3,-0.5) node[below] {$width(R)$};

\draw[->] (3.8,3.1) -- (3.4,2.5);
\draw (6.0,3.4) node {$p = Area_R = \P [\, \text{picked } \theta \text{ in } R]\, $};
\draw[->] (0.0,-0.5) node[left] {Good $\theta$} -- (1.5,-0.2);

\end{tikzpicture}
} 

&

\resizebox{4.5cm}{!}{
\begin{tikzpicture}
\def\normaltwo{\x,{4*1/exp(((\x-3)^2)/(2*0.5))}}

\newcommand{\xiValue}[1]{%
4*1/exp(((#1-3)^2)/(2*0.5))
}

\fill [fill=orange!60] (1.6,0) -- plot[domain=1.6:4.4] (\normaltwo) -- (4.4,0) -- cycle;

\draw[color=blue,domain=0:6] plot (\normaltwo) node[right] {};

\draw[dashed] (4.4,{\xiValue{4.4}}) -- (4.4,0);
\draw[dashed] (1.6,{\xiValue{1.6}}) -- (1.6,0);

\draw (0,5) node[above] {$\text{PMF}^{\text{Boosted}}(\theta)$};
\draw (6.2,0) node[right] {$\theta \in \Theta$};

\draw[->] (0,0) -- (6.2,0) node[right] {};
\draw[->] (0,0) -- (0,5) node[above] {};
\filldraw[fill=green, draw=green!40!black] (1.6,0.3) rectangle (4.4,-0.2);
\draw (3,0) node {Region $R$};

\draw[color=white] (3,-0.5) node[below] {$.$};

\end{tikzpicture}
} 
\\
Before $\A_{\text{Boost}}$ & After $\A_{\text{Boost}}$ \\
\end{tabular}
\end{center}
\end{table}

\[ p = Area_R = \P [\, \text{picked } \theta \text{ in } R]\,\]

While we have certain probability $p$ when we do the measure on $| \theta \rangle$, we want to make the process more efficiently by using the amplitude amplification in the Grover's search manner. In this paper we simply call this process ``Boosting'', as it boosts the PMF from $\text{PMF}(\theta)$ to $\text{PMF}^{\text{Boosted}}$.

We then construct the algorithm $\A_{\text{Boost}}$ by using the standard logic of amplitude amplification according to section 7.3 of \cite{wolf2019quantum}. We use the oracle gate $O_{flag}$ and the $\RR$ gate (i.e. ``keeping zero flipping else'') to construct the reflection $| B \rangle$ and $| U \rangle$:

\[ \begin{cases} O_{flag}: | \theta \rangle \mapsto (-1)^{flag(\theta)} | \theta \rangle \\
\RR = \begin{bmatrix}
1 & & & &\\
 & -1 & & &\\
 & & -1 & &\\
 & & & \ddots &\\
 & & & & -1\\
\end{bmatrix} \end{cases}\]

\[ \begin{cases}
ref(|B\rangle) = O_{flag} \\
ref(|U\rangle) = \A_{\text{1-query}} \RR \A^{-1}_{\text{1-query}}
\end{cases} \]

Under the classic approach, the original number of $\A_{\text{1-query}}$ required to get a good $\theta \in R$ is roughly $\frac{1}{p} = 1/\P[\, \text{picked } \theta \text{ in } R \text{ under } \A_{\text{1-query}}]\,$. By doing amplitude amplification, we can reduce the number of inquiries from $O(\frac{1}{p})$ to $O(\frac{1}{\sqrt{p}})$. Therefore we are improving from ``$\P \Big[ \text{picked } \theta \text{ in } R \text{ by } \#(\A_{\text{1-query}})=O(\frac{1}{p}) \text{ via classic algorithm} \Big] \approx 1$'' to ``$\P \Big[ \text{picked } \theta \text{ in } R \text{ by } \#(\A_{\text{1-query}})=O(\frac{1}{\sqrt{p}}) \text{ via } \A_{\text{Boost}} \Big] = \P \Big[ \text{picked } \theta \text{ in } R \text{ under } \A_{\text{Boost}} \Big] \approx 1$''.

\begin{algorithm}
  \caption{$\A_{\text{Boost}}:$ To amplify the probability of finding a good $\theta$}
  \begin{algorithmic}[1]
      \State \% \textbf{Main idea}: To boost $\P[\, \text{picked } \theta \text{ in region } R ]\,$ by looping $\A_{\text{1-query}}$
      \State \% \textbf{Input}: $\begin{cases} \Theta, \quad \xi_{threshold}, \quad \ell_{threshold} \\ \{ \sum\limits_{\theta \in \Theta} | E_{\theta j} \rangle : x_j \in S\} \\ \sum\limits_{\theta \in \Theta} | \theta \rangle \end{cases}$
      \State \% \textbf{Sub-algorithm}: $\A_{\text{1-query}}$  \Comment{Complexity of $\A_{\text{1-query}}: O(N)$}
      \State \% \textbf{Output}: $\theta_{\text{boost}}$
      \State \% \textbf{Analysis 1}: $\P[\, \A_{\text{1-query}} \text{ picked } \theta \text{ in region } R ]\, = p$
      \State \% \textbf{Analysis 2}: $\P[\, \A_{\text{Boost}} \text{ picked } \theta \text{ in region } R ]\, \approx 1$
      \State \% \textbf{Complexity}: $O(\frac{N}{\sqrt{Area_R}}) = O(\frac{N}{\sqrt{p}}) \leq O(\sqrt{|\Theta|} N)$
      \State Initial state: $| U \rangle = \A_{\text{1-query}} | \textbf{0} \rangle$
      \For{$k = 1:O(\frac{1}{\sqrt{p}})$} \Comment{Amplitude amplification boosting: $O(\frac{1}{p}) \mapsto O(\frac{1}{\sqrt{p}})$}
      \State $ref(| B \rangle)$
      \State $ref(| U \rangle)$
      \EndFor
      \State \textbf{Return} $\theta_{\text{Boost}} = $ \Qcircuit @C=0em @R=2em {\lstick{} & \meter} $\; | \theta \rangle$ if $flag(\theta_{\text{Boost}}) = 1$ \Comment{(i.e. Good $\theta$ in region $R$)}
  \end{algorithmic}
\end{algorithm}

Please note that the amplitude amplification works no matter the subroutine $\A_{\text{1-query}}$ is a classic algorithm or quantum algorithm, as per section 7.3 of \cite{wolf2019quantum}. Besides, $| \textbf{0} \rangle = | 0^m \rangle$, where $m$ is the number of qubits where $|\Theta| = 2^m$. For $| U \rangle = \A_{\text{1-query}} | \textbf{0} \rangle$ and $ref(|U\rangle) = \A_{\text{1-query}} \RR \A^{-1}_{\text{1-query}}$, here we are using a slight abuse of notation, while the actual meaning of $\A_{\text{1-query}}$ here is $Gate_{\A_{\text{1-query}}}$ since $\A_{\text{1-query}}$ is only an algorithm, but it can uniquely determine a quantum gate $Gate_{\A_{\text{1-query}}}$.

\subsubsection{Potential Cost Improvement}
\label{section:potential_cost_improvement}

The cost of $\A_{\text{Boost}}$ is $O(\frac{N}{\sqrt{p}})$. According to Lemma 2.5 (proof: section \ref{section:lemma_2_5} in Appendix), $O(\frac{N}{\sqrt{p}}) \leq O(\sqrt{|\Theta|} N)$. As mentioned in the proof, the inequality is ``very loose'', which should have much room to be improved.

It is worth to point out that the cost $O(N)$ of $\A_{\text{Boost}}$ originates from $\A_{\text{1-query}}$. In $\A_{\text{1-query}}$, the cost for quantum parallelism is $O(1)$ only, while the cost $O(N)$ is mainly due to the checking process of whether $\theta \in R$ by running the classic summation $\A_\xi$.

As we will see, the overall cost of this approach 2 (i.e. $\A_{\text{PSTC}}$; the cut-off approach PSTC) is $O(\sqrt{|\Theta|} N)$, which at the first glance is no better than approach 1. But the cost $O(N)$ in approach 1 is fundamental (i.e. built in the quantum parallelism), while the cost $O(N)$ in approach 2 originates from the checking process. Therefore it is very probable that the cost can be improved further to $O(sublinear \ N)$ by doing enhancement on the checking process frequency (e.g. dynamic frequency or sampling), as well as using clever dynamic choice of $\xi_{threshold}$.

We will also point out that successfully reducing cost from $O(|\Theta|)$ to $O(\sqrt{|\Theta|})$ is much more important than reducing cost from $O(N)$ to $O(sublinear \ N)$ in view of the fundamental property of the data collection process of type II uniformity. For details, please refer to Section \ref{section:why_O_N_doesnt_matter} of the Appendix.

\subsection{$\A_{\text{PSTC}}$: The final main algorithm for approach 2 (PSTC)}

Finally, to complete approach 2 (i.e. the cut-off approach PSTC), the main algorithm $\A_{\text{PSTC}}$ is to be implemented.

\begin{algorithm}
  \caption{$\A_{\text{PSTC}}:$ the main algorithm}
  \begin{algorithmic}[1]
      \State \% \textbf{Main idea}: Main algorithm to find $\theta$ with low loss
      \State \% \textbf{Input}: $\Theta$
      \State \% \textbf{Sub-algorithm}: $\A_{\text{Boost}}$ \Comment{Complexity of $\A_{\text{Boost}}: O(\sqrt{|\Theta|} N)$}
      \State \% \textbf{Output}: $\theta_{\text{main}}$
      \State \% \textbf{Complexity}: $O(\sqrt{|\Theta|} N)$ \Comment{Not just $O(\sqrt{|\Theta|} N \log |\Theta|)$}
      \State $| \textbf{0} \rangle \xmapsto[]{H_\Theta} \frac{1}{\sqrt{|\Theta|}} \sum\limits_{\theta \in \Theta} | \theta \rangle$
      \State $| \textbf{0} \rangle \xmapsto[]{H_S} \frac{1}{\sqrt{N}} \sum\limits_{x_j \in S} | x_j \rangle = \frac{1}{\sqrt{N}} \sum\limits_{j = 1}^N | x_j \rangle$
      \State Randomly sample $\theta_\text{Best} \in \Theta$ and $ \tilde{\theta} \triangleq \theta_{threshold} \in \Theta$
      \State $\ell_{threshold} \mapsfrom \frac{1}{N} \sum\limits_j \ell(\tilde{\theta}, x_j)$
      \State $\xi_{threshold} \mapsfrom \A_\xi(\tilde{\theta}, \ell_{threshold})$
      \State $\LL_\text{Best} \mapsfrom \frac{1}{N} \sum\limits_{j=1}^N \One[\, \ell(\theta_{\text{Best}}, x_j) \leq \ell_{threshold} ]\,$
      \For{$k = 1:O(\log |\Theta|)$}
      \State $\sum\limits_{\theta \in \Theta} \sum\limits_{j = 1}^N | \theta \rangle | x_j \rangle | \textbf{0} \rangle \xmapsto[]{\text{Quantum Parallelism}} \sum\limits_{\theta \in \Theta} \sum\limits_{j = 1}^N | \theta \rangle | x_j \rangle | E_{\theta, j} \rangle$
      \State $\theta_{\text{Boost}} \mapsfrom \A_{\text{Boost}}$ \Comment{ Parameters $\sum\limits_{\theta} \theta, \sum\limits_{\theta, j} | E_{\theta, j} \rangle$ etc. will be passed into $\A_{\text{Boost}}$}
      \State $\LL_\text{Boost} \mapsfrom \frac{1}{N} \sum\limits_{j=1}^N \One[\, \ell(\theta_{\text{Boost}}, x_j) \leq \ell_{threshold} ]\,$
		\If{$\LL_\text{Boost} \geq \LL_\text{Best} $}
		\State $\theta_\text{Best} \mapsfrom \theta_{\text{Boost}}$ \Comment{Update $\theta_\text{Best}$ if $\LL$ is further maximized}
		\EndIf
      \EndFor
      \State \textbf{Return} $\theta_{\text{main}} = \theta_\text{Best}$
  \end{algorithmic}
\end{algorithm}

In $\A_{\text{PSTC}}$, we are updating ``better'' $\theta$ who is defined as having a larger $\LL^{\text{PSTC}}_\theta$ than the current $\theta_{\text{Best}}$. In each update of $\theta$, we are roughly updating the objective function by half, and therefore we requires a for-loop of $O(\log |\Theta|)$ times.

By taking the $\A_{\text{Boost}}$ with cost $O(\sqrt{|\Theta|} N)$ for $O(\log |\Theta|)$ times, it appears that the final cost would be $O(\sqrt{|\Theta|} N \log |\Theta|)$ instead of $O(\sqrt{|\Theta|} N)$. But don't forget that region $R$ in later iteration is narrower than the previous iterations, and that size roughly reducing by half in the new iteration. Therefore finally our cost is:

\[ Cost_{\text{Final iteration}} + ... + Cost_{\text{1st iteration}} = O(\sqrt{|\Theta|} + \frac{\sqrt{|\Theta|}}{2} + \frac{\sqrt{|\Theta|}}{4} + ...) = O(\sqrt{|\Theta|}) \]


\section{Equivalency discussion between $\LL^{avg}$ and $\LL^{\text{PSTC}}$}
\label{section:equivalency_discussion}

Note here that the parameters $\{ \theta \}$ themselves are not integers. However, they are drawn from a set of size $M$ (usually $M=2^m$ for some $m$) and so we have indexed them by $i\in \{1,2,..M\}$. Also, for any $i$, the parameter $\theta_i$ indexed by $i$ is smaller than the one indexed by $i+1$. So, the distance between the indices ($i$) mirrors that between the parameters indexed ($\theta_i$). Since in approach 1 and approach 2 we are using different objective function $\theta_{avg}^{*}$ and $\theta_{PSTC}^{*}$ in (\ref{eq:quantum_parallelism_avg}) and (\ref{eq:quantum_parallelism_PSTC}), it is also worth to discuss whether this two different objective function are `` equivalent'' in some sense.
Equivalency here means that there are some constants $C,c$ independent of the cardinality of the set of parameters and the sample set such that on any sample set,

 \[c\theta_{PSTC}^{*} \leq \theta_{avg}^{*} \leq C \theta_{PSTC}^{*}.\]
 
The above inequality is referring to each dimension of $\theta$ for $dim(\theta) \geq 1$, and we assume that all $\theta$ are restricted to be positive (this setting is good enough because we are going to prove that $\theta^{*}_{\text{PSTC}}$ and $\theta^{*}_{avg}$ are not equivalent). We now construct a simple counter example with $dim(\theta) = 1$ to disprove the above inequality. Suppose such constants $c,C$ existed. Note that without loss of generality, we can assume $C+1$ is a power of $2$.  Consider a sample set of size 3 as below: $C+1=M$. Let $\ell_{threshold}=2$ and suppose that for all $i\in {2,...C }$, we have the losses: \[ \ell_{i,1}=2.5, \ell_{i,2}=\ell_{i,3}=2\] 
 and suppose that,
 
 \[\ell_{1,1}= \ell_{1,2}= \ell_{1,3}=1.9\]\[  \ell_{C+1,1}= 2.1, \ell_{C+1,2}=1, \ell_{C+1,3}=0 \] 
 
Here,  $\theta_{avg}^{*}= \argmin\limits_{\theta} \frac{1}{3} \sum\limits_{j=1}^3 \ell_{\theta j} = C+1$, but $\theta_{PSTC}^{*}=\argmax\limits_{\theta} \frac{1}{3} \sum\limits_{j=1}^3 \One [\, \ell_{\theta j} \leq \ell_{threshold} ]\, =1$ . So, $C+1 \leq C$ gives a contradiction. 

\begin{figure}[ht]
    \centering
    \includegraphics[width=0.50\textwidth]{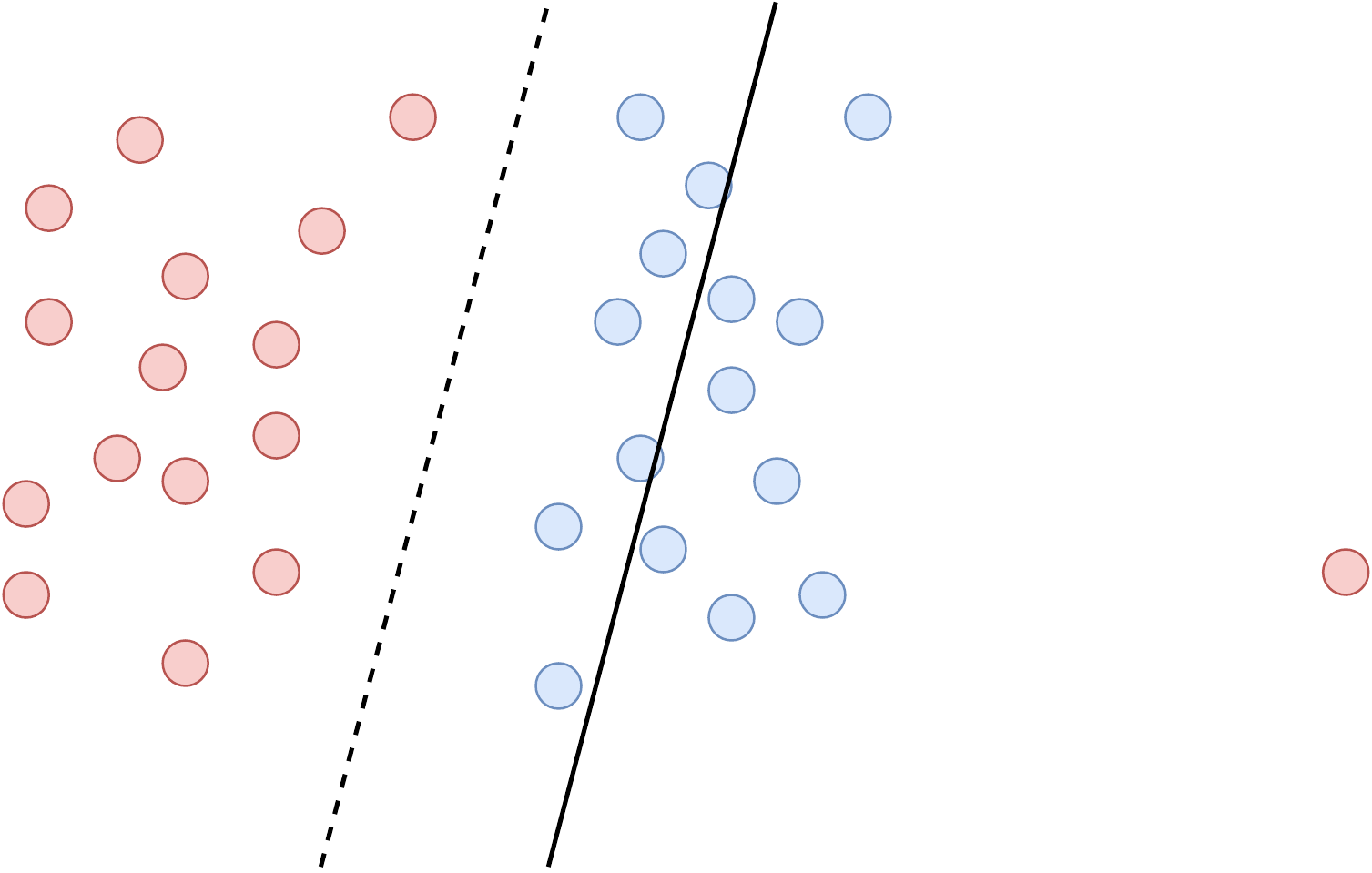}
    \caption{An example of outliers in linear classification}
    \label{fig:noise_example}
\end{figure}

The intuition behind the estimator $\theta_{PSTC}^{*}$ for the modified objective is that estimator $\theta_{avg}^{*}$ for the traditional loss function is less robust, i.e. $\theta_{avg}^{*}$ is more easily influenced by noise and outliers, which are quite common in a large training dataset. In the above example, the outlier $\ell_{C+1, 3}=0$ causes $\theta_{avg}^{*}=C+1$, but $\theta_{PSTC}^{*}=1$ is not impacted by $\ell_{C+1, 3}$. Figure~\ref{fig:noise_example} shows an example of linear classification. The rightmost red dot is apparently an outlier and the black solid line is a possible classification boundary obtained from traditional objective, while the black dashed line is a more reasonable boundary given this data distribution. Our hope is that the modified objective will be able to exclude the impacts of these outliers from our model and generate a more robust model.

\section{Conclusion}
\label{section:conclusion}

We had proposed two approaches to optimize the learning model, namely the average approach and the cut-off approach (PSTC), which is based on a novel objective function (\ref{eq:quantum_parallelism_PSTC}) by applying the cut-off indicator $E_{\theta j}$.

As mentioned in section \ref{section:potential_cost_improvement}
 (Potential Cost Improvement), currently the two approaches have the same cost $O(\sqrt{|\Theta|} N)$. Since $\theta$ is the parameter to be searched, while $N$ is referring to the number of samples, it is not expected that the Grover search would reduce the cost $O(N)$ and therefore $O(\sqrt{|\Theta|} N)$ is an acceptable level of cost. We also mentioned in \ref{section:why_O_N_doesnt_matter} (Appendix) that reducing cost from $O(N)$ to $O(sublinear \ N)$ is much less important than reducing cost from $O(|\Theta|)$ to $O(\sqrt{|\Theta|})$ in view of the fundamental property of the data collection process of type II uniformity.
 
Meanwhile, we showed that the cost $O(N)$ of average approach is mainly at the quantum parallelism part, and the $O(N)$ of PSTC is mainly at the checking process, and the cost of each quantum parallelism in PSTC is $O(1)$ only. Therefore there is potential to arithmetically improve PSTC to reduce cost from $O(\sqrt{|\Theta|} N)$ to $O(\sqrt{|\Theta|} \cdot sublinear \ N)$ in future work.

\section{Future Work and Extensions}
\label{section:future_work_extension}

One future work and two extensions are suggested. Future work would enhance the current framework, while the two extensions extend the application areas of the current framework.

\subsection{Future work: Cost reduction for PSTC}

As mentioned in the section \ref{section:potential_cost_improvement}, the current cost of average approach and cut-off approach (PSTC) are same as $O(\sqrt{|\Theta|} N)$, but it is very probable that the cost of PSTC can be improved further to $O(\sqrt{|\Theta|} \cdot sublinear \ N)$ by doing enhancement on the checking process frequency (e.g. dynamic frequency or sampling), as well as using clever dynamic choice of $\xi_{threshold}$.

Such algorithmic enhancement will be left as a future work.

\subsection{Extension: Additional Restriction}

There is another benefit of extending our framework from approach 1 (i.e. the average approach) to approach 2 (i.e. the cut-off approach PSTC).

In optimization problem, we would normally search $\theta$ over $\Theta$ as below:

\[ \argmin\limits_{\theta \in \Theta} \frac{1}{N} \sum\limits_{j = 1}^N \ell(\theta, x_j) \]

It is also common for us to have additional condition on $\theta$ (e.g. fairness of model on some special features say gender/race \cite{Woodworth2017}, required smoothness of model, or the overall parameter size $\| \theta \|_2$ restricted by regularization etc. ). Therefore our problem may be re-formulated as below:

\[ \argmin\limits_{\theta \in \Theta \cap A} \frac{1}{N} \sum\limits_{j = 1}^N \ell(\theta, x_j) \]

where $A$ refers to some additional restriction/condition on $\theta$. In view of this, we can generalize the cut-off indicator $E_{\theta j}$ of (\ref{eq:L_cut_off}) to $E^A_{\theta j}$ as below:

\begin{equation}
\label{eq:additional_condition_A}
\LL^{A, \text{PSTC}}_\theta \triangleq \frac{1}{N} \sum\limits_{j=1}^N E^A_{\theta j} \text{, where } E^A_{\theta j} = \One[\, \ell(\theta, x_j) \leq \tilde{\ell} \text{ and } \theta \in A]\,
\end{equation}

And then we can optimize $\LL^{A, \text{PSTC}}_\theta$ in the same manner that we optimized $\LL^{\text{PSTC}}_\theta$ in the cut-off approach. We have a simple example scenario of using $E^A_{\theta j}$ in section \ref{section:adv_example}. Note that the additional condition $A$ won't significantly increase the complexity of the problem, while it often do under the classic setting. 

\subsection{Extension: Adversarial Example}
\label{section:adv_example}

In machine learning, the topic of Adversarial example \cite{Adv_Example} is getting much attention from the researchers in the last 10 years. In short, the problem of adversarial example is to seek a perturbed input $x^{adv}$ (with change $\delta^{adv}$ unperceivable by human) from the original $x$, such that $x^{adv}$ will be misclassified by our target model $h_\theta$.

\begin{figure}[h!]
\caption{Adversarial example}
\centering
\includegraphics[scale=0.4]{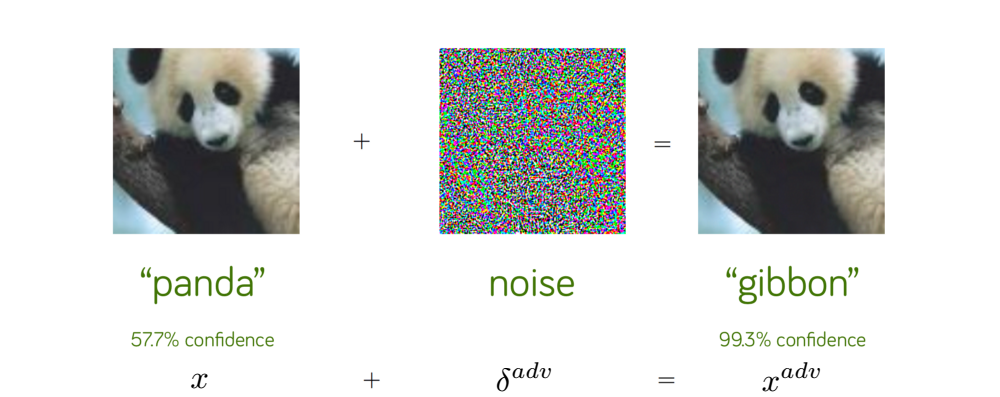}
\end{figure}

One version of the adversarial problem can be formulated in this way:

\begin{equation}
\label{eq:adv_example}
x^{adv} = x + \delta^{adv} = x + \argmax\limits_{\| \delta \| \leq \epsilon} \ell (h_\theta (x + \delta), y)
\end{equation}

Note that we are maximizing the loss $\ell$ since we want to fool the model $h_\theta$. In this problem, the role of $x$ and $\theta$ are switched: $\theta$ is now fixed, and $x$ is the parameters to be optimized.

To solve (\ref{eq:adv_example}), we can simply apply PSTC by searching for ``good'' $\delta \in X$ instead of $\theta \in \Theta$ in all places that require parameter. We now have an additional condition ``$\| \delta \| \leq \epsilon$'', but shouldn't be a problem as we can generalize our cut-off-qubit from $E_{\theta j}$ to $E^{\| \delta \| \leq \epsilon}_{\theta j}$ in the manner of (\ref{eq:additional_condition_A}). After that, we will be able to obtain an adversarial example in the cost $O(\sqrt{|X|})$.

In adversarial example, the searching process for perturbation $\delta$ refers to type I uniformity (see section \ref{section:why_O_N_doesnt_matter}), and therefore we talk about $|X|$ instead of $N = |S|$. But it's fine as the cost is now reduced from $O(|X|)$ to $O(\sqrt{|X|})$.

\pagebreak
\section{Notations table}
\label{section:notations_table}

\begin{table}[ht]
\resizebox{15cm}{!}{%
\begin{tabular}{| c | c | c |}

\hline
Notation & Space & Meaning\\

\hline
$\theta$ & $\Theta$ & Parameter of model $h_\theta(\cdot)$\\
$h_\theta(\cdot)$ & $X \rightarrow Y$ & Model with parameter $\theta$\\
$x$ & $X$ & An input\\
$y$ & $Y$ & A label\\
$S$ & $S \subseteq X$ & $\{x_1,..., x_N\}$\\
$N$ & $\Z^{+}$ & $\#(\text{samples in } S)$\\
$n$ & $\Z^{+}$ & $N = 2^n$ \\
$m$ & $\Z^{+}$ & $|\Theta| = 2^m$ \\
\hline
$\ell(\cdot, \cdot)$, or $loss(\cdot, \cdot)$ & $\Theta \times X \rightarrow \R^{+}$ & Loss function\\
& (or $Y \times Y \rightarrow \R^{+}$) & (depending on context)\\
$\ell_{ij}$, or $\ell_{\theta_i j}$, or $\ell_{\theta j}$ & $\R^{+}$ & $\ell_{\theta_i j} = \ell(\theta_i, x_j) = \ell(h_{\theta_i}(x_j), y_j)$ \\
$\LL^{avg}_{\theta}$ or $\LL^{avg}(\theta)$ & $\R^{+}$  & Average loss $\frac{1}{N} \sum\limits_{j=1}^N \ell_{\theta j}$; The smaller the better\\
$\LL^{\text{PSTC}}_{\theta}$ or $\LL^{\text{PSTC}}(\theta)$ & $\R^{+}$ & Cut-off loss $\frac{1}{N} \sum\limits_{j=1}^N \One[\, \ell_{\theta j} \leq \tilde{\ell} ]\,$; The larger the better\\
$\tilde{\ell}$, or $\ell_{threshold}$ & $\R^{+}$ & Threshold value for the cut-off approach\\
$E_{\theta j}$ & $\{ 0, 1\}$ & Cut-off indicator: Value $\One[\, \ell_{\theta j} \leq \tilde{\ell}]\,$ to be stored in 1 qubit\\
$| \phi \rangle, | \psi \rangle$ & $\C^{n + 1}$ & Pure states with meaning defined in (\ref{eq:phi_and_psi})\\
\hline
$ Q_{\text{1-query}}$ & Q-Circuits & The Q-circuit for partial swap test \\
$ | \varsigma_\theta \rangle $ & $\C^{2 n  + 3}$ & The pure state after the 2nd Hadamard gate in $ Q_{\text{1-query}}$ given $\theta$\\
$| \varsigma^{(0)}_\theta \rangle, | \varsigma^{(1)}_\theta \rangle$ & $\C^{2 n  + 2}$ & The $| 0 \rangle$ and $| 1 \rangle$ parts of $ | \varsigma_\theta \rangle $\\
$\text{PMF}(\cdot)$ & $\Theta \rightarrow [\, 0, 1]\, $ & Probability mass function (PMF) of $\theta$\\
$\A_{\text{1-query}}$, $\A_{\xi}$, $\A_{\text{Boost}}$, $\A_{\text{PSTC}}$ & Algorithms & Names of some cut-off approach algorithms\\
$\theta_{\text{1-query}}$, $\theta_{\text{Boost}}$, $\theta_{\text{Best}}$ & $\Theta$ & $\theta$ being used in the respective algorithms\\
$flag(\cdot)$ & $\Theta \rightarrow \{ 0, 1\}$ & Flag to indicate whether $\theta$ is good or bad\\
$O_{\text{flag}}$ & $\C^{m \times m}$ & Gate of oracles on $flag(\theta)$\\
$\RR$ & $\C^{m \times m}$ & Gate of ``keeping zero, flipping else''\\
$R$ & $R \subseteq \Theta$ & Region $R$ that ``good'' $\theta$s concentrate\\
$ref(|B\rangle)$ & $\C^{m \times m}$ & Reflection on the ``Bad'' vector in $\A_{\text{Boost}}$\\
$ref(|U\rangle)$ & $\C^{m \times m}$ & Reflection on the ``Uniform'' vector in $\A_{\text{Boost}}$\\
$p$ & $[\, 0, 1]\, \in \R$ & $\P[\, \text{picked } \theta \text{ in } R]\,$\\
$\xi(\cdot)$ & $\Theta \rightarrow \R^{+}$ & A function that proportional to pdf of $\theta: P[\, \text{observe } \theta | \text{1st qubit} = 0 ]\,$\\
$\tilde{\xi}$, or $\xi_{threshold}$ & $\R^{+}$ & Threshold value of $\xi$ for region $R$\\
$| \textbf{0} \rangle$ & $\C^{\text{not care}}$ & High dimensional $| 0 \rangle$ with dimension not mentioned\\
\hline
$x^{adv}$ & X & Adversarial example\\
$\delta^{adv}$ & X & Perturbation\\
$|\chi^{\text{Type I}}\rangle$, $|\chi^{\text{Type II}}\rangle$ & $\C^{\text{depends}}$ & Pure states of $x$ referring to Type I and Type II uniformity.\\
\hline

\end{tabular}
} 
\end{table}

$\linebreak$
$\linebreak$
$\linebreak$
$\linebreak$
$\linebreak$
$\linebreak$
$\linebreak$
$\linebreak$
$\linebreak$
$\linebreak$
$\linebreak$
$\linebreak$
$\linebreak$
$\linebreak$
$\linebreak$
$\linebreak$
$\linebreak$
$\linebreak$

\pagebreak
\section{Appendix}
\label{section:appendix}

\subsection{Lemma 2.0}
\label{section:lemma_2_0}
In $Q_{\text{1-query}}$ (i.e. Figure \ref{fig:Q_1_query}),

\begin{equation}
\label{eq:lemma_2_0_pre}
| \varsigma_\theta \rangle \triangleq | 0 \rangle | \varsigma^{(0)}_\theta \rangle + | 1 \rangle | \varsigma^{(1)}_\theta \rangle = \frac{1}{2} | 0 \rangle \Big( | \phi_\theta \rangle | \psi \rangle + | \psi \rangle | \phi_\theta \rangle \Big) + \frac{1}{2} | 1 \rangle \Big( | \phi_\theta \rangle | \psi \rangle - | \psi \rangle | \phi_\theta \rangle \Big)
\end{equation}

\begin{equation}
\label{eq:lemma_2_0}
\frac{1}{\sqrt{|\Theta|}} \sum\limits_{\theta \in \Theta} | \varsigma_\theta \rangle | \theta \rangle = \frac{1}{\sqrt{|\Theta|}} \sum\limits_{\theta \in \Theta} \frac{1}{2} | 0 \rangle \Big( | \phi_\theta \rangle | \psi \rangle + | \psi \rangle | \phi_\theta \rangle \Big) | \theta \rangle + \frac{1}{2} | 1 \rangle \Big( | \phi_\theta \rangle | \psi \rangle - | \psi \rangle | \phi_\theta \rangle \Big) | \theta \rangle
\end{equation}

\begin{proof}
$ | \varsigma_\theta \rangle $ is the $(2 n + 3)$-qubit-state (i.e. $1+2 \cdot (n+1)$) after the 2nd Hadamard gate in $ Q_{\text{1-query}}$ for a given $\theta$. The standard result of swap test \cite{Swap_Test} is (\ref{eq:lemma_2_0_pre}).

If we also included the last qubit of $\theta$ in the state, the overall $(2 n + 3 + m)$-qubit state would be (\ref{eq:lemma_2_0}).
\end{proof}

\subsection{Lemma 2.1}
\label{section:lemma_2_1}

\begin{equation}
\label{eq:lemma_2_1}
\P \Big[ \text{1st qubit} = 0 | \theta \Big] = \frac{1}{2} + \frac{1}{2} |\langle \phi_\theta, \psi \rangle |^2
\end{equation}

\begin{proof}

\[ (\ref{eq:lemma_2_0_pre}) \implies \P \Big[ \text{1st qubit} = 0 | \theta \Big] = \langle \varsigma^{(0)}_\theta | \varsigma^{(0)}_\theta \rangle = \frac{1}{2} \Big( \langle \phi_\theta | \langle \psi | + \langle \psi | \langle \phi_\theta | \Big)\frac{1}{2} \Big( | \phi_\theta \rangle | \psi \rangle + | \psi \rangle | \phi_\theta \rangle \Big)\]

\[ = \frac{1}{4} \Big( 2 \langle \phi_\theta | \phi_\theta \rangle \langle \psi | \psi \rangle + \langle \psi | \phi_\theta \rangle \langle \phi_\theta | \psi \rangle + \langle \phi_\theta | \psi \rangle \langle \psi | \phi_\theta \rangle \Big) = \frac{1}{2} + \frac{1}{2} |\langle \phi_\theta, \psi \rangle |^2 \]

\end{proof}

\subsection{Lemma 2.2}
\label{section:lemma_2_2}

\begin{equation}
\label{eq:lemma_2_2}
\P \Big[ \text{1st qubit} = 0 \Big] = \frac{1}{2} + \frac{1}{2 | \Theta |} \sum\limits_{\theta \in \Theta} |\langle \phi_\theta, \psi \rangle |^2
\end{equation}

\begin{proof}

\[ \P \Big[ \text{1st qubit} = 0 \Big] = \Big( \frac{1}{\sqrt{|\Theta|}} \sum\limits_{\theta \in \Theta} \langle \varsigma^{(0)}_\theta | \langle \theta | \Big)  \Big( \frac{1}{\sqrt{|\Theta|}} \sum\limits_{\theta \in \Theta} | \varsigma^{(0)}_\theta \rangle | \theta \rangle \Big) 
\]

\[ = 
(\ref{eq:lemma_2_0}) \implies \frac{1}{\sqrt{| \Theta |}} \sum\limits_{\alpha \in \Theta} \frac{1}{2} \Big( \langle \phi_\alpha | \langle \psi | + \langle \psi | \langle \phi_\alpha | \Big) \langle \alpha | \cdot \frac{1}{\sqrt{| \Theta |}} \sum\limits_{\beta \in \Theta} \frac{1}{2} \Big( | \phi_\beta \rangle | \psi \rangle + | \psi \rangle | \phi_\beta \rangle \Big) | \beta \rangle \]

\[ = \frac{1}{4 | \Theta |} \sum\limits_{\alpha, \beta \in \Theta} \Big( \langle \phi_\alpha | \phi_\beta \rangle \langle \psi | \psi \rangle + \langle \psi | \phi_\beta \rangle \langle \phi_\alpha | \psi \rangle +  \langle \phi_\alpha | \psi \rangle \langle \psi | \phi_\beta \rangle + \langle \psi | \psi \rangle \langle \phi_\alpha | \phi_\beta \rangle \Big) \langle \alpha | \beta \rangle \]

\[ = \frac{1}{4 | \Theta |} \sum\limits_{\alpha, \beta \in \Theta} \Big( 2 \langle \phi_\alpha | \phi_\beta \rangle + \langle \psi | \phi_\beta \rangle \langle \phi_\alpha | \psi \rangle +  \langle \phi_\alpha | \psi \rangle \langle \psi | \phi_\beta \rangle \Big) \delta_{\alpha \beta} \text{ (i.e. } \delta \text{ is Kronecker delta}) \]

\[ = \frac{1}{4 | \Theta |} \sum\limits_{\theta \in \Theta} \Big( 2 \langle \phi_\theta | \phi_\theta \rangle + 2 |\langle \phi_\theta, \psi \rangle |^2 \Big) \cdot 1 = \frac{1}{2} + \frac{1}{2 | \Theta |} \sum\limits_{\theta \in \Theta} |\langle \phi_\theta, \psi \rangle |^2 \]

\end{proof}

Alternative proof: We can prove (\ref{eq:lemma_2_2}) by $\P \Big[ \text{1st qubit} = 0 \Big] = \frac{1}{|\Theta|} \P \Big[ \text{1st qubit} = 0 \Big | \text{observe } \theta \Big]$ if we prove (\ref{eq:lemma_2_3}) first instead.

Note: (\ref{eq:lemma_2_2}) also tell us that $\P \Big[ \text{1st qubit} \Big]$ is roughly $O(1)$ given that our choice of $\tilde{\ell} \triangleq \ell_{threshold}$ is proper.

\subsection{Lemma 2.3}
\label{section:lemma_2_3}

\begin{equation}
\label{eq:lemma_2_3}
\P \Big[ \text{observe } \theta \Big] = \frac{1}{|\Theta|}
\end{equation}

\begin{proof}

\[ (\ref{eq:lemma_2_0}) \implies \P \Big[ \text{observe } \theta \Big] = \Big( \frac{1}{\sqrt{|\Theta|}} \langle \varsigma_\theta | \Big) \Big( \frac{1}{\sqrt{|\Theta|}} | \varsigma_\theta \rangle \Big) = \frac{1}{|\Theta|}\Big( \langle 0 | \langle \varsigma^{(0)}_\theta | + \langle 1 | \langle \varsigma^{(1)}_\theta | \Big) \Big( | 0 \rangle | \varsigma^{(0)}_\theta \rangle + | 1 \rangle | \varsigma^{(1)}_\theta \rangle \Big) \]

\[ = \frac{1}{|\Theta|} \Big( \langle 0 | 0 \rangle \langle \varsigma^{(0)}_\theta | \varsigma^{(0)}_\theta \rangle + \langle 1 | 1 \rangle \langle \varsigma^{(1)}_\theta | \varsigma^{(1)}_\theta \rangle \Big) = \frac{1}{|\Theta|} \Big( \frac{1}{2} + \frac{1}{2} |\langle \phi_\theta, \psi \rangle |^2 + \frac{1}{2} - \frac{1}{2} |\langle \phi_\theta, \psi \rangle |^2\Big) = \frac{1}{|\Theta|}\]
\end{proof}

\subsection{Theorem 2.4}
\label{section:theorem_2_4}

\begin{equation}
\label{eq:theorem_2_4}
\P \Big[ \text{observe } \theta | \text{1st qubit = 0} \Big] \propto \frac{1}{2} + \frac{1}{2} \Big| \frac{1}{N} \sum\limits_{j = 1}^N E_{\theta j} \Big|^2
\end{equation}

\begin{proof}

According to Bayes Rules, we have:

\[ \P[\, \text{observe } \theta | \text{1st qubit} = 0]\, = \frac{\P[\, \text{1st qubit} = 0 | \text{observe } \theta]\, \cdot \P [\, \text{observe } \theta ]\,}{\P[\, \text{1st qubit} = 0]\,} \]

\[ = \frac{ \Big( \frac{1}{2} + \frac{1}{2} |\langle \phi_\theta, \psi \rangle |^2 \Big) \cdot \frac{1}{|\Theta|}}{ \frac{1}{2} + \frac{1}{2 | \Theta |} \sum\limits_{\theta \in \Theta} |\langle \phi_\theta, \psi \rangle |^2} \text{ (i.e. by } (\ref{eq:lemma_2_1}), (\ref{eq:lemma_2_2}), (\ref{eq:lemma_2_3})) \]

\[ \propto \frac{1}{2} + \frac{1}{2} |\langle \phi_\theta, \psi \rangle |^2 \text{ (i.e. removing all multiples unrelated to } \theta ) \]

\[ = \frac{1}{2} + \frac{1}{2} \Big| \frac{1}{N} \sum\limits_{j = 1}^N E_{\theta j} \Big|^2 \text{(i.e. by (\ref{eq:max_L_cut_off}))} \]

\end{proof}

\subsection{Lemma 2.5}
\label{section:lemma_2_5}

\begin{equation}
\label{eq:lemma_2_5}
O(\frac{1}{\sqrt{p}}) \leq O(\sqrt{|\Theta|})
\end{equation}

\begin{proof}

In the figures of table \ref{table:PMF_xi_theta}, the $Area_R$ and $width(R)$ is now formally defined as below:

\[ \begin{cases} Area_R \triangleq P[\, \text{picked } \theta \text{ in } R ]\, \propto \sum\limits_{\theta \in R} \xi(\theta) \\ width(R) \triangleq |R| \end{cases}\]

\[ O(\frac{1}{\sqrt{p}}) = O(\frac{1}{\sqrt{Area_R}}) = O(\frac{1}{\sqrt{\P [\, \text{picked } \theta \text{ in } R \text{ under } \A_{\text{1-query}}]\,}}) \]

\[ =  O(\frac{1}{\sqrt{ \sum\limits_{i \in R} \P [\, \theta_i | \text{1st qubit in } \A_{\text{1-query}} = 0 ]\,}}) = O(\frac{1}{\sqrt{\sum\limits_{\theta_i \in R} \xi(\theta_i)}}) \]

\[ \Big( \text{Note: } (\ref{eq:lemma_2_2}) \implies \P [\, \text{1st qubit in } \A_{\text{1-query}} = 0 ]\, \approx O(1) \Big) \]

\[ = O(\frac{1}{\sqrt{\sum\limits_{\theta_i \in R} \frac{1}{2} + \frac{1}{2} \Big| \frac{1}{N} \sum\limits_{j = 1}^N E_{\theta_i j} \Big|^2}}) = O(\frac{1}{\sqrt{\sum\limits_{\theta_i \in R} \Big| \frac{1}{N} \sum\limits_{j = 1}^N E_{\theta_i j} \Big|^2}})\]

\[ = O(\frac{1}{\sqrt{\sum\limits_{\theta_i \in R} 1}}) = O(\frac{1}{\sqrt{width(R)}}) \leq O(\sqrt{|\Theta|})\]

\end{proof}

Note that the last inequality sign is ``very loose'', which have much room to be improved.

The potential improvement depends on how ``concentrated'' we want for region $R$, i.e. the choice of $\xi_{threshold}$. If we don't require concentrated region, the cost $O(\frac{1}{\sqrt{width(R)}})$ can get much closer to the $O(1)$ side than the $O(\sqrt{|\Theta|})$ side. But in that case the machine learning problem optimization may not be very meaningful as we want $\theta$ to be optimized to a certain extent. So that will be a trade-off between goal and cost, and in any case we have the bound $O(\sqrt{|\Theta|})$.

\subsection{Why reducing cost from $O(N)$ can be less important than what it appears to be?}
\label{section:why_O_N_doesnt_matter}

In our approach 2 (i.e. the cut-off approach PSTC), we have shown that the cost of $\A_{\text{1-query}}$ is $O(N)$, while the overall cost of $\A_{\text{PSTC}}$ is $O(N \sqrt{|\Theta})$. As a smaller cost is always welcome, it is reasonable to pursue the cost reduction further, and as mentioned in Section \ref{section:potential_cost_improvement}, there is potential for future work to reduce the cost of $\A_{\text{Boost}}$ further by doing arithmetic enhancement on the check process or the $\xi_{threshold}$ selection.

In spite of the above, it is also worth to point out that reducing cost from $O(N)$ to $O(sublinear \ N)$ can be much less important than reducing cost from $O(|\Theta|)$ to $O(\sqrt{|\Theta|})$.

\subsubsection{Two types of uniformity}

Taking computer vision as example. We take the feature space as $X = \R^{H \times W \times 3}$ (i.e. height/width/color of photos). There are 2 types of possible uniform distribution of $x \in X: $

\[ \text{Uniform distribution: } \begin{cases} \text{Type I uniformity (real-world-distribution): } |\chi^{\text{Type I}}\rangle = \frac{1}{constant}\sum\limits_{x \in \R^{H \times W \times 3}} | x \rangle \\ \text{Type II uniformity (entire-space-uniform-distribution): } |\chi^{\text{Type II}}\rangle = \frac{1}{constant}\sum\limits_{x \in S} | x \rangle \end{cases} \]

where $S = \{ x_1, x_2, ..., x_N \}= \{ photo_1, photo_2, ..., photo_N \} \subseteq \R^{H \times W \times 3}$. Note that the number of samples of type I uniformity is much larger than type II uniformity. Therefore much more qubits is required to store $|\chi^{\text{Type II}}\rangle$ than $|\chi^{\text{Type I}}\rangle$. For most cases in machine training, we only cares about type II uniformity, since we are collecting data under the ``real-world-distribution'' in the manner of Figure \ref{fig:type_II_uniformity} instead of the entire-space-uniform-distribution:

\begin{figure}[h!]
\caption{Type II uniformity in computer vision data collection process}
\label{fig:type_II_uniformity}
\centering
\includegraphics[scale=0.8]{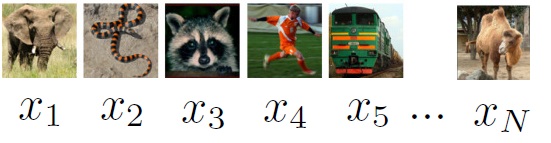}
\end{figure}

For most of the cases of the data collection process of type II uniformity, the data are collected in the economical cost of $O(N)$ (e.g. salary of the photographers/providers). Therefore even if the computation cost at the algorithm is cheaper than $O(N)$, the overall cost is still $O(N)$ in view of the economical cost of data collection. 

As a result, our focus for the cost should be the $O(\sqrt{|\Theta|})$ part of $O(\sqrt{|\Theta|} N)$, since the $O(N)$ part is unavoidable at the side of the type II data collection.

\subsubsection{Purpose and paradigm of research}

Another reason why the cost $O(N)$ may be less important than we expected is that in many deep learning research, researchers are interested in testing different models with different architecture/parameters to achieve specific computational goal, while using the same set of data. Therefore it's the hypothesis class (with different size of $|\Theta|$), as well as the parameter $\theta$ being changed, that matters. In contrast, in most of the cases, the number of samples $N$ won't change much for a specific well defined task.

In view of the above, successfully reducing the cost from $O(|\Theta|)$ to $O(\sqrt{|\Theta|})$ is much more important than reducing the cost from $O(N)$ to $O(\text{sub-linear } N)$, as the cost $O(N)$ could be less expensive than what it appears to be, though we had suggested some potential ways in Section \ref{section:potential_cost_improvement} to further reduce the cost $O(N)$ to $O(sublinear \ N)$.

\medskip
\small
\bibliography{./8_ref} 

\end{document}